\documentclass[
 reprint,
 amssymb, amsmath,
 aip,jcp
]{revtex4-1}

\usepackage{comment} 
\usepackage{dcolumn} 
\usepackage[colorlinks=true,linkcolor=blue]{hyperref}%
\expandafter\ifx\csname package@font\endcsname\relax\else
 \expandafter\expandafter
 \expandafter\usepackage
 \expandafter\expandafter
 \expandafter{\csname package@font\endcsname}%
\fi
\usepackage{xspace}

\usepackage{graphicx}
\usepackage{amsmath}

\usepackage[normalem]{ulem}
\usepackage[utf8]{inputenc}

\newcommand{\eexact}[0]{E_0}
\newcommand{\ebasis}[0]{E_0^{\basis}}

\newcommand{\pbeontns}{\text{PBE-OT}}                                                                                  
\newcommand{\argrpbeontns}[0]{\den(\br{}),\nabla n(\br{}),\ntwoextrapcas(\br{}),\mu_{\text{}}^{}(\br{})}                       
\newcommand{\ntwoextrapcas}[0]{\mathring{n}_2^{\text{}}}
\newcommand{\ntwoextrap}[0]{\mathring{n}_{2}}

\newcommand{\argecmd}[0]{\den,\nabla n,\ntwo,\mu}

\newcommand{\ket}[1]{{\ensuremath{|#1\rangle}\xspace}}
\newcommand{\bra}[1]{{\ensuremath{\langle #1|}\xspace}}
\newcommand{\elemm}[3]{\bra{#1} #2 \ket{#3}}

\newcommand{\basis}[0]{\mathcal{B}}

\newcommand{\den}[0]{{n}}

\newcommand{\denz}[0]{{n_0}}

\newcommand{\denrbasis}[0]{{n}^{\basis}({\bf r})}

\newcommand{\denzrbasis}[0]{{n_0^\basis}({\bf r})}
\newcommand{\denzbasis}[0]{{n_0^\basis}}
\newcommand{\denbasis}[0]{{n}^{\basis}}

\newcommand{\denpsibasis}[0]{\den_{\psibasis}}

\newcommand{\psibasis}[0]{\Psi^{\basis}}

\newcommand{\psifci}[0]{\Psi^{\basis}_{\text{FCI}}}

\newcommand{\kinop}[0]{\hat{T}}

\newcommand{\weeop}[0]{\hat{W}_{\text{ee}}}

\newcommand{\efci}[0]{E_{\text{FCI}}^{\basis}}

\newcommand{\efuncbasis}[0]{\bar{E}^\basis[\denbasis]}
\newcommand{\efuncbasispsi}[0]{\bar{E}^\basis[n_{\psibasis},n_{2,\!\psibasis}]}

\newcommand{\psizbasis}[0]{\Psi_0^\basis}

\newcommand{\pbeuegXi}{\text{PBE-UEG}}

\newcommand{\levelcomp}[2]{\text{CIPSI+#1}_{#2}}

\newcommand{\br}[1]{{\mathbf{r}_{#1}}}                                                                                                       
\newcommand{\dbr}[1]{d\br{#1}}

\newcommand{\brb}[1]{{\bf r}_{#1}}
\newcommand{\rab}[1]{|\brb{1} - \brb{2}|}

\newcommand{\ai}[1]{\hat{a}_{#1}}
\newcommand{\aic}[1]{\hat{a}^{\dagger}_{#1}}

\newcommand{\argrpbeuegXi}[0]{\den(\br{}),\nabla n(\br{}),\ntwo^{\text{UEG}}(\br{}),\mu_{\text{}}(\br{})}

\newcommand{\twodmrdiagpsi}[0]{ n_{2,\Psi^\basis_{\text{loc}}}(\br{})}

\newcommand{\fbasis}[0]{f^{\basis}_{\Psi^\basis_{\text{loc}}}(\br{})}

\newcommand{\wbasiscoal}[1]{W^{\basis}(\br{})}

\newcommand{\SO}[2]{\phi_{#1}(\br{#2})}

\newcommand{\murhf}[0]{\mu_{\text{SD}}}

\newcommand{\murfci}[0]{\mu_{\text{CIPSI}}}

\newcommand{\fbasisval}[0]{f^{\basis}_{\Psi^\basis_{\text{loc,val}}}(\br{})}
\newcommand{\twodmrdiagpsival}[0]{n_{2,\Psi^\basis_{\text{loc,val}}}(\br{})}
\newcommand{\argepbe}[0]{\den,\nabla n}

\newcommand{\ecmd}[0]{\bar{e}_{\text{c,md}}^{\text{sr},\text{PBE}}}                                                                                                                               
\newcommand{\ntwo}[0]{n_{2}}

\newcommand{\vbar}[0]{\hat{\bar{V}}^\basis[\denbasis]}

\newcommand{\PT}[1]{\scriptstyle[#1]}

\begin{document}

\author{Emmanuel Giner}%
\email{emmanuel.giner@lct.jussieu.fr}
\affiliation{Laboratoire de Chimie Théorique, Sorbonne Université and CNRS, F-75005 Paris, France}
\author{Diata Traore}
\affiliation{Laboratoire de Chimie Théorique, Sorbonne Université and CNRS, F-75005 Paris, France}
\author{Barth\' elemy Pradines}
\affiliation{Laboratoire de Chimie Théorique, Sorbonne Université and CNRS, F-75005 Paris, France}
\affiliation{Institut des Sciences du Calcul et des Données, Sorbonne Université, F-75005 Paris, France}
\author{Julien Toulouse}
\email{toulouse@lct.jussieu.fr}
\affiliation{Laboratoire de Chimie Théorique, Sorbonne Université and CNRS, F-75005 Paris, France}
\affiliation{Institut Universitaire de France, F-75005 Paris, France}

\title{Self-consistent density-based basis-set correction: How much do we lower total energies and improve dipole moments?} 

\begin{abstract}
This work provides a self-consistent extension of the recently proposed density-based basis-set correction method for wave-function electronic-structure calculations [J. Chem. Phys. {\bf 149}, 194301 (2018)]. In contrast to the previously used approximation where the basis-set correction density functional was a posteriori added to the energy from a wave-function calculation, here the energy minimization is performed including the basis-set correction. Compared to the non-self-consistent approximation, this allows one to lower the total energy and change the wave function under the effect of the basis-set correction. This work addresses two main questions: i) What is the change in total energy compared to the non-self-consistent approximation, and ii) can we obtain better properties, namely dipole moments, with the basis-set corrected wave functions? We implement the present formalism with two different basis-set correction functionals and test it on different molecular systems. The main results of the study are that i) the total energy lowering obtained by the self-consistent approach is extremely small, which justifies the use of the non-self-consistent approximation, and ii) the dipole moments obtained from the basis-set corrected wave functions are improved, being already close to their complete-basis-set values with triple-zeta basis sets. Thus, the present study further confirms the soundness of the density-based basis-set correction scheme.
\end{abstract}

\date{March 25, 2021}
\maketitle

\section{Introduction}

One of the main limitations of electronic-structure calculations based on wave-function theory (WFT) is the slow convergence of the results with respect to the size of the one-electron basis set. This aspect is of fundamental importance in quantum chemistry as WFT methods have otherwise many interesting features. In particular, in a given basis set $\basis$, WFT methods can usually be systematically improved toward the exact solution provided by full configuration interaction (FCI). At the root of the slow convergence of WFT lies the singularity of the Coulomb interaction: because it becomes infinite at electron-electron coalescence points, it creates a derivative discontinuity of the wave function at these points, the so-called electron-electron cusp\cite{Hyl-ZP-29,Kat-CPAM-57}, which is not representable by the usual incomplete basis sets employed. To cure the slow convergence problem of WFT, two main approaches have emerged: basis-set extrapolation techniques~\cite{HelKloKocNog-JCP-97,HalHelJorKloKocOlsWil-CPL-98} and explicitly correlated F12 methods\cite{Ten-TCA-12,TenNog-WIREs-12,HatKloKohTew-CR-12, KonBisVal-CR-12, GruHirOhnTen-JCP-17, MaWer-WIREs-18}. Basis-set extrapolation techniques consist in exploiting the known asymptotic behavior of WFT properties as a function of the size of the basis set in order to estimate the complete-basis-set (CBS) limit based on several calculations with basis sets of increasing sizes. Explicitly correlated F12 methods consist in supplementing to the usual basis sets a correlation factor which restores the electron-electron cusp and accelerates the convergence toward the CBS limit. Although these F12 methods are increasingly popular in quantum chemistry\cite{TewKloNeiHat-PCCP-07}, they have the drawback of needing rather complex three- and four-electron integrals\cite{BarLoo-JCP-17} and, more generally, of involving a relatively complex mathematical formalism which makes the adaptation of a WFT method to its F12 version a non trivial task.

An alternative path to speed up the convergence of WFT calculations with respect to the size of the basis set has been recently proposed by some of the present authors\cite{GinPraFerAssSavTou-JCP-18} by exploiting the ability of range-separated density-functional theory (RSDFT) to recover the short-range correlation effects missing from an incomplete basis set. The central idea developed in Ref.~\onlinecite{GinPraFerAssSavTou-JCP-18} is to define a mapping between the electron-electron Coulomb interaction projected into an incomplete basis set $\basis$ and the non-diverging long-range electron-electron interaction $\text{erf}(\mu r_{12})/r_{12}$ used in RSDFT. The connection is done through the definition of a range-separation parameter $\mu$ which varies in space and automatically adapts to the basis set $\basis$. Once this adaptive range-separation parameter is defined, one can use a special flavor of short-range correlation density functionals used in RSDFT for the estimation of the correlation energy missing in the considered basis set $\basis$. An important property of this RSDFT-based approach is that the basis-set energy correction properly vanishes in the CBS limit. This strategy was successfully  validated for the calculations of ionization potentials\cite{GinPraFerAssSavTou-JCP-18,LooPraSceGinTou-JCTC-20}, molecular atomization energies\cite{LooPraSceTouGin-JCPL-19,GinSceLooTou-JCP-20,YaoGinLiTouUmr-JCP-21}, and excitation energies\cite{GinSceTouLoo-JCP-19}. 

All the previous applications of this method rely on a non-self-consistent approximation in which the basis-set energy correction is just added a posteriori to a good estimate of the FCI energy in a given basis set $\basis$. In the present work, we go beyond this approximation and develop a self-consistent formalism in order to answer two distinct questions: i) How crude is the non-self-consistent approximation for total energies; and ii) can the self-consistent formalism yield effective wave functions with better properties? 

The paper is organized as follows. In Section \ref{sec:exact_theo} we present the exact theory of the self-consistent basis-set correction scheme using either a functional of the density only or a functional of both the density and the on-top pair density and we recall the non-self-consistent approximation previously employed. In Section \ref{sec:approx_efunc} we introduce our approximations of the unknown exact basis-set correction functional by short-range functionals. In Section \ref{sec:cipsi} we explain how we solve the self-consistent basis-set correction equations by a selected configuration interaction algorithm. In Section \ref{sec:numerical}, we report and discuss results on the total energies of the Be atom and the BH molecule, and on the dipole moments of the BH, FH, H$_2$O, and CH$_2$ molecules. Section~\ref{sec:conclusion} contains our conclusions. Unless otherwise specified, Hartree (Ha) atomic units (a.u.) are used throughout the paper.

\section{Theory}
\label{sec:theory}

\subsection{Self-consistent basis-set correction}
\label{sec:exact_theo}

\subsubsection{Basis-set correction as a functional of the density}
\label{sec:EBasfuncdens}

We start by reviewing the scheme where the basis-set correction is written as a functional of the density, which was first developed in Ref.~\onlinecite{GinPraFerAssSavTou-JCP-18}. Given a basis set $\basis$, the exact ground-state energy $\eexact$ of an electronic system can be approximated by the energy $\ebasis$ obtained by the following minimization over $\basis$-representable one-electron densities $\denbasis$, i.e. densities that can be obtained from a wave function $\psibasis$ belonging to the $N$-electron Hilbert space generated by the basis set $\basis$,
\begin{equation}
 \label{eq:E_b}
 \ebasis = \min_{\denbasis} \left\{  F[\denbasis] + \int \text{d}{\bf r} \,\,v_{\text{ne}}({\bf r}) \, \denrbasis
\right\},
\end{equation}
where $v_{\text{ne}}({\bf r})$ is the nuclei-electron interaction potential. In this expression, $F[\denbasis]$ is the standard Levy-Lieb constrained-search universal density functional~\cite{Lev-PNAS-79,Lie-IJQC-83} evaluated at $\denbasis$
\begin{equation}
 \label{eq:F_lieb}
 F[\denbasis] = \min_{\Psi \rightarrow \denbasis} \elemm{\Psi}{\kinop + \weeop}{\Psi},
\end{equation}
where $\kinop$ and $\weeop$ are the kinetic-energy operator and the Coulomb electron-electron interaction operator, respectively, and the notation $\Psi \rightarrow \denbasis$ means a $N$-electron wave function yielding the density $\denbasis$. It is important to notice that the wave functions $\Psi$ used in the definition of $F[\denbasis]$ in Eq.~\eqref{eq:F_lieb} are not restricted to be expandable in the basis set $\basis$ but should instead be thought of as expanded on a complete basis set. The minimizing density $\denzbasis$ in Eq.~\eqref{eq:E_b} can be considered as the best variational approximation to the exact ground-state density $\denz$. Importantly, when the basis set $\basis$ is increased up to the CBS limit, the density $\denzbasis$ and energy $\ebasis$ converge to the exact ground-state density $\denz$ and energy $\eexact$, respectively,
\begin{equation}
 \begin{aligned}
 & \lim_{\basis \rightarrow \text{CBS}} \denzbasis = \denz \;\;\; \text{and} & \lim_{\basis \rightarrow \text{CBS}} \ebasis = \eexact. 
 \end{aligned}
\end{equation}
Since the $\basis$-representability restriction is only applied to densities and not to wave functions, the basis-set convergence of $\ebasis$ to $\eexact$ is much faster than in a usual WFT calculation.

We then decompose the universal Levy-Lieb density functional as 
\begin{equation}
\label{eq:F_lieb2}
 F[\denbasis] = \min_{\psibasis \rightarrow \denbasis}  \elemm{\psibasis}{\kinop + \weeop}{\psibasis} +  \efuncbasis, 
\end{equation}
where $\psibasis$ designates wave functions restricted to the $N$-electron Hilbert space generated by the basis set $\basis$, and $\efuncbasis$ is the complementary basis-set correction density functional required to make Eq.~\eqref{eq:F_lieb2} exact.
This basis-set correction functional $\efuncbasis$ recovers the part of the energy that is missing in the first term of the right-hand side of Eq.~\eqref{eq:F_lieb2} due to the basis-set restriction of the wave functions $\psibasis$. 
Inserting Eq.~\eqref{eq:F_lieb2} into Eq.~\eqref{eq:E_b} and recombining the two minimizations, we can obtain $\ebasis$ by the following minimization over $\basis$-restricted wave functions $\psibasis$
\begin{equation}
 \label{eq:E_minpsi}
  \begin{aligned}
 \ebasis = \min_{\psibasis} \Big\{\elemm{\psibasis}{\kinop + \weeop +  \hat{V}_\text{ne}}{\psibasis} + \bar{E}^\basis[n_{\psibasis}]\Big\},
  \end{aligned}
\end{equation}
where $\hat{V}_\text{ne}$ is the nuclei-electron interaction operator and $\denpsibasis(\br{})=\langle \psibasis |\hat{n}(\br{})| \psibasis \rangle $ is the density of the wave function $\psibasis$, where we have introduced the density operator $\hat{n}(\br{}) = \sum_{\sigma \in \{\uparrow,\downarrow\}}  \hat{\psi}_\sigma^\dagger(\br{}) \hat{\psi}_\sigma(\br{})$ written in real-space second quantization. The minimizing wave function $\Psi_0^\basis$ in Eq.~\eqref{eq:E_minpsi} satisfies the following self-consistent Schr\"odinger-like equation
\begin{equation}
 \label{eq:HeffB}
  \begin{aligned}
 \hat{H}_\text{eff}^\basis[n_{\Psi_{0}^\basis}] \ket{\psizbasis} = {\cal E}_0^\basis \ket{\psizbasis},
  \end{aligned}
\end{equation}
where ${\cal E}_0^\basis$ is the Lagrange multiplier associated with the normalization constraint of the wave function $\psizbasis$ and the effective Hamiltonian is defined for a given density $\denbasis$ as
\begin{equation}
 \label{eq:def_HeffB}
  \begin{aligned}
  \hat{H}_\text{eff}^\basis[\denbasis] =  \hat{T}^\basis + \weeop^\basis + \hat{V}_\text{ne}^\basis + \vbar.
  \end{aligned}
\end{equation}
In this expression, $\hat{T}^\basis$, $\weeop^\basis$, and $\hat{V}_\text{ne}^\basis$ are the kinetic, electron-electron, and electron-nuclei operators projected in the $N$-electron Hilbert space generated by the basis set $\basis$, and $\vbar$ is the one-electron effective potential operator
\begin{equation}
 \label{eq:def_vb}
 \vbar= \int \text{d}{\bf r} \; \bar{v}^\basis({\bf r}) \,\, \hat{n}^\basis({\bf r}),
\end{equation}
where $\bar{v}^\basis({\bf r}) = \delta \bar{E}^\basis[\denbasis]/\delta \denrbasis$ and $\hat{n}^\basis({\bf r})$ is the density operator projected in the basis set $\basis$.

Using real-valued spatial orthonormal orbitals $\{ \phi_p \}$ spanning the same space as the basis set $\basis$, the expression of the effective Hamiltonian in second quantization is
\begin{equation}
 \label{eq:Heffsq}
 \begin{aligned}
 \hat{H}_\text{eff}^\basis[\denbasis] = \sum_{pq}^\basis \big( h_{pq} + \bar{v}_{pq}^\basis \big) \hat{E}_{pq}  
                  + \frac{1}{2} \sum_{pqrs}^\basis w_{pqrs} \hat{e}_{pqrs},
 \end{aligned}
\end{equation}
where $\hat{E}_{pq} = \hat{a}^\dagger_{p\uparrow} \hat{a}_{q\uparrow} + \hat{a}^\dagger_{p\downarrow} \hat{a}_{q\downarrow}$ and $\hat{e}_{pqrs} = \hat{E}_{pr} \hat{E}_{qs} - \delta_{qr} \hat{E}_{ps}$ are the spin-singlet one- and two-particle elementary operators, $h_{pq}$ are the usual one-electron integrals, $w_{pqrs} = \langle pq|rs \rangle$ are the usual two-electron integrals, and $\bar{v}_{pq}^\basis$ are the one-electron integrals associated with the effective potential $\bar{v}^\basis({\bf r})$
\begin{equation}
 \label{eq:v_ij}
 \bar{v}_{pq}^\basis = \int \text{d} \br{} \phi_{p}(\br{}) \bar{v}^\basis({\bf r})  \phi_q(\br{}). 
\end{equation}
Note that in Eq.~\eqref{eq:Heffsq}, we put $\basis$ on top of the sum symbols to indicate that the sums run over all orbitals generated by the basis set $\basis$. 

Finally, note that we have considered the total density $n^\basis$ for simplicity but the  theory can be trivially extended to spin densities $n_\uparrow^\basis$ and  $n_\downarrow^\basis$.

\subsubsection{Basis-set correction as a functional of the density and the on-top pair density}
\label{sec:EBasfuncdensandot}

We now extend the theory to allow for a basis-set correction functional depending on both the density $n_{\Psi^\basis}(\br{})$ and the on-top pair density $n_{2,\Psi^\basis}(\br{})=\langle \psibasis |\hat{n}_2(\br{})| \psibasis \rangle$ of a wave function $\Psi^\basis$, where we have introduced the on-top pair density operator $\hat{n}_2(\br{}) = \sum_{\sigma \in \{\uparrow,\downarrow\}} \sum_{\sigma' \in \{\uparrow,\downarrow\}} \hat{\psi}_\sigma^\dagger(\br{}) \hat{\psi}_{\sigma'}^\dagger(\br{}) \hat{\psi}_{\sigma'}(\br{}) \hat{\psi}_\sigma(\br{})$. In the spirit of the generalized Kohn-Sham scheme~\cite{SeiGorVogMajLev-PRB-96} (see also Ref.~\onlinecite{Tou-INC-21}), we write the universal Levy-Lieb density functional as
\begin{equation}
\label{eq:F_lieb3}
 F[\denbasis] = \min_{\psibasis \rightarrow \denbasis}  \left\{ \elemm{\psibasis}{\kinop + \weeop}{\psibasis} +  \efuncbasispsi \right\}, 
\end{equation}
where $\efuncbasispsi$ can be any functional of $n_{\Psi^\basis}$ and $n_{2,\Psi^\basis}$ such that the minimization in Eq.~(\ref{eq:F_lieb3}) exactly gives $F[\denbasis]$. Insertion into Eq.~\eqref{eq:E_b} leads to 
\begin{equation}
 \label{eq:E_minpsi2}
  \begin{aligned}
 \ebasis = \min_{\psibasis} \Big\{\elemm{\psibasis}{\kinop + \weeop +  \hat{V}_\text{ne}}{\psibasis} + \efuncbasispsi \Big\},
  \end{aligned}
\end{equation}
and the minimizing wave function $\Psi_0^\basis$ satisfies the following self-consistent Schr\"odinger-like equation
\begin{equation}
 \label{eq:HeffB2}
  \begin{aligned}
 \hat{H}_\text{eff}^\basis[n_{\Psi_{0}^\basis},n_{2,\Psi_{0}^\basis}] \ket{\psizbasis} = {\cal E}_0^\basis \ket{\psizbasis},
  \end{aligned}
\end{equation}
where the effective Hamiltonian is defined for a given $\basis$-representable density $\denbasis$ and on-top pair density $n_2^\basis$ as
\begin{equation}
 \label{eq:def_HeffB2}
  \begin{aligned}
  \hat{H}_\text{eff}^\basis[n^\basis,n_2^\basis] =  \hat{T}^\basis + \weeop^\basis + \hat{V}_\text{ne}^\basis + \hat{\bar{V}}^\basis[n^\basis,n_2^\basis] 
\\
+ \hat{\bar{W}}^\basis[n^\basis,n_2^\basis].
  \end{aligned}
\end{equation}
In this expression, $\hat{\bar{V}}^\basis[n^\basis,n_2^\basis]$ is the one-electron effective potential operator
\begin{equation}
 \label{eq:def_vb2}
 \hat{\bar{V}}^\basis[n^\basis,n_2^\basis] = \int \text{d}{\bf r} \; \bar{v}^\basis({\bf r}) \,\, \hat{n}^\basis({\bf r}),
\end{equation}
where $\bar{v}^\basis({\bf r}) = \delta \bar{E}^\basis[n^\basis,n_2^\basis]/\delta \denrbasis$, and $\hat{\bar{W}}^\basis[n^\basis,n_2^\basis]$ is the two-electron effective interaction operator
\begin{equation}
 \label{eq:def_wb2}
 \hat{\bar{W}}^\basis[n^\basis,n_2^\basis] = \frac{1}{2} \int \text{d}{\bf r} \; \bar{w}^\basis({\bf r}) \,\, \hat{n}_2^\basis({\bf r}),
\end{equation}
where $\bar{w}^\basis({\bf r}) = 2 \delta \bar{E}^\basis[n^\basis,n_2^\basis]/\delta n_2^\basis({\bf r})$ and $\hat{n}_2^\basis({\bf r})$ is the on-top pair density operator projected in the basis set $\basis$. The second-quantized expression of the effective Hamiltonian is
\begin{equation}
 \begin{aligned}
 \hat{H}_\text{eff}^\basis[n^\basis,n_2^\basis] =& \sum_{pq}^\basis \big( h_{pq} + \bar{v}_{pq}^\basis \big) \hat{E}_{pq}  
\\
          & + \frac{1}{2} \sum_{pqrs}^\basis \left( w_{pqrs} + \bar{w}_{pqrs}^\basis \right) \hat{e}_{pqrs},
 \end{aligned}
\end{equation}
where, as before, $\bar{v}_{pq}^\basis$ are the one-electron integrals associated with the effective potential $\bar{v}^\basis({\bf r})$, and $\bar{w}_{pqrs}^\basis$ are the two-electron integrals associated with the effective interaction $\bar{w}^\basis({\bf r})$
\begin{equation}
 \label{eq:w_ijkl}
 \bar{w}_{pqrs}^\basis = \int \text{d} \br{} \phi_{p}(\br{})\phi_{r}(\br{}) \bar{w}^\basis({\bf r})  \phi_q(\br{}) \phi_s(\br{}). 
\end{equation}

Of course, since the effective Hamiltonians in Eqs.~\eqref{eq:def_HeffB} and~\eqref{eq:def_HeffB2} are different, their respective ground-state wave function $\Psi_0^\basis$ are also different, even though we used the same notation.

\subsubsection{Non-self-consistent approximation}
\label{sec:nonself}

In previous works~\cite{GinPraFerAssSavTou-JCP-18,LooPraSceTouGin-JCPL-19,GinSceTouLoo-JCP-19,YaoGinLiTouUmr-JCP-21}, the minimization in Eq.~\eqref{eq:E_minpsi} or ~\eqref{eq:E_minpsi2} was not performed but the minimizing wave function $\Psi_0^\basis$ was simply approximated by the standard FCI wave function $\Psi_\text{FCI}^\basis$ (or an estimate of it) in the basis set $\basis$
\begin{equation}
 \label{eq:psi0}
  \psizbasis \approx  \psifci, 
\end{equation}
leading to the following approximation for $\ebasis$, for the basic theory of Section~\ref{sec:EBasfuncdens},
\begin{equation}
 \label{eq:E_B_fci}
  \ebasis \approx   \efci +  \bar{E}^\basis[n_{\psifci}], 
\end{equation}
where $\efci$ is the standard FCI energy (or an estimate of it) in the basis set $\basis$, and for the extended theory of Section~\ref{sec:EBasfuncdensandot}
\begin{equation}
 \label{eq:E_B_fci2}
  \ebasis \approx   \efci +  \bar{E}^\basis[n_{\psifci},n_{2,\psifci}]. 
\end{equation}
The approximation in Eq.~\eqref{eq:psi0} is in fact equivalent to approximating the minimizing density $n_0^\basis$ in Eq.~\eqref{eq:E_b} by the standard FCI ground-state density $n_{\psifci}$
\begin{equation}
 \label{eq:n0_eq_nfci}
 \denzrbasis \approx  n_{\psifci}({\bf r}), 
\end{equation}
which seems intuitively a reasonable approximation as one expects $n_{\psifci}$ and $n_0^\basis$ to be both close to the exact density $n_0$, and the encouraging numerical results obtained for energies with this non-self-consistent approximation tend to confirm the validity of Eq.~\eqref{eq:n0_eq_nfci}. Nevertheless, in the present study, we will investigate the quantitative effect on energies and dipole moments of performing the minimization in Eq.~\eqref{eq:E_minpsi} or~\eqref{eq:E_minpsi2}.

\subsection{Approximations for the basis-set correction functional $\bar{E}^\basis$}
\label{sec:approx_efunc}

\subsubsection{Local range-separation parameter}
\label{sec:mur}

As originally proposed in Ref.~\onlinecite{GinPraFerAssSavTou-JCP-18}, the basis-set correction functional $\bar{E}^\basis$ can be mapped to the so-called short-range correlation functional with multideterminant reference introduced in Ref.~\onlinecite{TouGorSav-TCA-05} in the context of RSDFT. This mapping relies on the definition of a local range-separation parameter $\mu^\basis(\br{})$~\cite{GinPraFerAssSavTou-JCP-18}
\begin{equation}
 \label{eq:def_mur} 
 \mu^\basis(\br{}) = \frac{\sqrt{\pi}}{2} \wbasiscoal, 
\end{equation}
which provides a local measure of the incompleteness of the basis set $\basis$. It is defined such that the long-range electron-electron interaction of RSDFT, $w^\text{lr}(r_{12})= \text{erf}(\mu r_{12})/r_{12}$, coincides at coalescence (i.e., at $r_{12}=0$) with an effective interaction representing the Coulomb electron-electron interaction projected in the basis set $\basis$. The expression of this effective interaction at coalescence is~\cite{GinPraFerAssSavTou-JCP-18}
\begin{equation}
 \label{eq:wbasis}
 \wbasiscoal\, =
    \begin{cases}
      \frac{\fbasis}{\twodmrdiagpsi},    & \text{if $\twodmrdiagpsi \ne 0$,}
\\
       \infty,                                                                                          & \text{otherwise,}
    \end{cases}
\end{equation}
with
\begin{equation}
        \label{eq:fbasis}
        \fbasis
        = \sum_{pqrstu}^\basis  w_{pqrs} \Gamma_{rstu}\SO{p}{} \SO{q}{} \SO{t}{} \SO{u}{},
\end{equation}
where $\Gamma_{pqrs} = 2 \elemm{\Psi_\text{loc}^\basis}{ \aic{r_\downarrow}\aic{s_\uparrow}\ai{q_\uparrow}\ai{p_\downarrow}}{\Psi_\text{loc}^\basis}$ is the opposite-spin two-electron density matrix of some ``localizing'' wave function $\Psi_\text{loc}^{\basis}$, and $\twodmrdiagpsi$ is its associated on-top pair density
\begin{equation}
 \twodmrdiagpsi = \sum_{pqrs}^\basis \Gamma_{pqrs} \SO{p}{} \SO{q}{}  \SO{r}{} \SO{s}{}. 
\end{equation}
The wave function $\Psi_\text{loc}^{\basis}$ is only used to localize the effective interaction projected in the basis set $\basis$. The local range-separation parameter is very weakly dependent on this wave function $\Psi_\text{loc}^{\basis}$. It should be thought of as essentially dependent on the basis set $\basis$. Importantly, in the CBS limit the effective interaction goes to the Coulomb interaction which diverges at coalescence and consequently the local range-separation parameter goes to infinity
\begin{equation}
 \label{eq:cbs_mu}
  \lim_{\basis \to \text{CBS}} \mu^\basis(\br{}) = \infty,
\end{equation}
independently of $\Psi_\text{loc}^{\basis}$, which is fundamental to guarantee the correct behavior of the theory in the CBS limit.

\subsubsection{Approximate basis-set correction functionals from short-range functionals} 
\label{sec:ecmd_basis}

Approximations for the basis-set correction functional $\bar{E}^\basis$ are obtained by using the previously defined local range-separation parameter in short-range correlation functionals. Specifically, for the basis-set correction functional in Eq.~(\ref{eq:E_minpsi}) we use the so-called PBE-UEG basis-set correction functional (UEG stands for ``uniform electron gas'')~\cite{LooPraSceTouGin-JCPL-19}
\begin{equation}
 \begin{aligned}
        \label{eq:def_pbeueg_i}
&       \bar{E}^\basis_{\pbeuegXi}[n] \\
&       =  \int  \dbr{}  \,  \ecmd(\argrpbeuegXi),
 \end{aligned}
\end{equation}
and, for the basis-set correction functional in Eq.~(\ref{eq:E_minpsi2}) we use the so-called PBE-OT basis-set correction functional (OT stands for ``on-top'')~\cite{GinSceLooTou-JCP-20}
\begin{equation}
 \begin{aligned}
        \label{eq:def_pbeueg_iv}
&        \bar{E}^\basis_{\pbeontns}[n,n_2] \\
&  = \int  \dbr{} \,  \ecmd(\argrpbeontns),   
 \end{aligned}
\end{equation}
where we have dropped the superscript $\basis$ in the density, in the on-top pair density, and in the local range-separation parameter for simplicity. In these expressions, the short-range (sr) correlation energy density with multideterminant (md) reference $\ecmd(\argecmd)$ has the following generic form in terms of the density $n$, the density gradient $\nabla n$, the on-top pair density $\ntwo$, and the range-separation parameter $\mu$~\cite{FerGinTou-JCP-18}
\begin{equation}
 \begin{aligned}
 \label{eq:def_ecmdpbe}
&\ecmd(\argecmd) = \frac{e_{\text{c}}^{\text{PBE}}(\argepbe)}{1+ \beta(\argepbe,\ntwo) \; \mu^3},
 \end{aligned}
\end{equation}
\begin{equation}
 \begin{aligned}
 \label{eq:def_ecmdpbe2}
&\beta(\argepbe,\ntwo) = \frac{e_{\text{c}}^{\text{PBE}}(\argepbe)}{c \; \ntwo}, 
 \end{aligned}
\end{equation}
where $e_{\text{c}}^{\text{PBE}}(\argepbe)$ is the usual Perdew-Burke-Ernzerhof (PBE) correlation energy density~\cite{PerBurErn-PRL-96} and $c=(2\sqrt{\pi}(1 - \sqrt{2}))/3$.

The key difference between the PBE-UEG and PBE-OT functional is the on-top pair density used. The PBE-UEG functional uses $\ntwo^{\text{UEG}}(\br{})$ which is an estimate of the exact on-top pair density using a parametrization of the on-top pair density of the uniform electron gas (UEG) at density $n({\br{}})$
\begin{equation}
 \label{eq:def_n2ueg}
 \ntwo^{\text{UEG}}({\br{}}) = n({\br{}})^2 g_0(n({\br{}})),                                                        
\end{equation}
where the on-top pair-distribution function $g_0(n)$ is taken from Eq.~(46) of Ref.~\onlinecite{GorSav-PRA-06}. By contrast, the PBE-OT functional uses $\ntwoextrapcas(\br{})$ which is an estimate of the exact on-top pair density obtained from extrapolating the input on-top pair density $\ntwo({\br{}})$ of the wave function $\Psi^\basis$ to the limit $\mu\to\infty$ (see Ref.~\onlinecite{GorSav-PRA-06})
\begin{equation}
 \label{eq:def_n2extrap}
 \ntwoextrap({\br{}}) = \bigg( 1 + \frac{2}{\sqrt{\pi}\mu({\br{}})} \bigg)^{-1} \; \ntwo({\br{}}).
\end{equation}
As shown in Ref.~\onlinecite{GinSceLooTou-JCP-20}, the difference between the two flavors of on-top pair densities comes from the treatment of strong correlation. While $\ntwo^{\text{UEG}}(\br{})$ is a good approximation of the exact on-top pair density for weakly correlated situations, when strong correlation effects are present it fails to represent the large depletion of the exact on-top pair density and in this case $\ntwoextrap(\br{})$ provides a much better approximation of the exact on-top pair density.

The explicit expression of the one-electron effective potential associated with the PBE-UEG functional,
\begin{equation}
\bar{v}^\basis_\pbeuegXi(\br{}) = \frac{\delta \bar{E}^\basis_{\pbeuegXi}[n]}{\delta n(\br{})},
\end{equation}
was already given in a previous work~\cite{LooPraSceGinTou-JCTC-20}. The corresponding potential for the PBE-OT functional,
\begin{equation}
\bar{v}^\basis_\pbeontns(\br{}) = \frac{\delta \bar{E}^\basis_{\pbeontns}[n,n_2]}{\delta n(\br{})},
\end{equation}
has a very similar expression, with the simplification that the on-top pair density $n_{2}$ in Eq.~\eqref{eq:def_pbeueg_iv} is taken as independent of the density whereas $n_{2}^\text{UEG}$ in Eq.~\eqref{eq:def_pbeueg_i} is a function of the density. For the PBE-OT functional, we have in addition the two-electron effective interaction
\begin{equation}
\bar{w}^\basis_\pbeontns(\br{}) = \frac{\delta \bar{E}^\basis_{\pbeontns}[n,n_2]}{\delta n_2(\br{})}.
\end{equation}
Its explicit expression is
\begin{equation}
\begin{aligned}
\bar{w}^\basis_\pbeontns(\br{}) = \frac{\partial \ecmd}{\partial n_2} (\argrpbeontns) \\
\times \frac{\partial \ntwoextrap(\br{})}{\partial n_2(\br{})}
\end{aligned}
\end{equation}
where
\begin{equation}
\begin{aligned}
\frac{\partial \ntwoextrap(\br{})}{\partial n_2(\br{})} =  \bigg( 1 + \frac{2}{\sqrt{\pi}\mu({\br{}})} \bigg)^{-1},
\end{aligned}
\end{equation}
and
\begin{equation}
\begin{aligned}
\frac{\partial \ecmd(\argecmd)}{\partial n_2} = \frac{\ecmd(\argecmd)^2 \; \mu^3}{c \; (n_2)^2}.
\end{aligned}
\end{equation}
Note that the localizing wave function $\Psi_\text{loc}^{\basis}$ used in the definition of the local range-separation parameter (see Section~\ref{sec:mur}) is taken as fixed, i.e. we do not consider variations of $\Psi_\text{loc}^{\basis}$. Importantly, since the local range-separation parameter diverges in the CBS limit [Eq.~\eqref{eq:cbs_mu}] and since the correlation energy per particle $\ecmd(\argecmd)$ in Eq.~\eqref{eq:def_ecmdpbe} vanishes for $\mu\to\infty$, both basis-set correction functionals correctly vanish in the CBS limit
\begin{equation}
 \begin{aligned}
\lim_{\basis \to \text{CBS}} \bar{E}^\basis_{\pbeuegXi}[n] = 0,
 \end{aligned}
\end{equation}
\begin{equation}
 \begin{aligned}
\lim_{\basis \to \text{CBS}} \bar{E}^\basis_{\pbeontns}[n,n_2] = 0,
 \end{aligned}
\end{equation}
i.e., the CBS limit is unaltered by the correction.

\subsubsection{Frozen-core version of the basis-set correction functionals}
\label{sec:FC}

When the wave function $\Psi^\basis$ is calculated in the frozen-core approximation, we use the frozen-core version of the basis-set correction functionals introduced in Ref.~\onlinecite{LooPraSceTouGin-JCPL-19}. The basis-set correction functionals become
\begin{equation}
\bar{E}^\basis[n_{\Psi^\basis}] \to \bar{E}^\basis[n_{\Psi^\basis_\text{val}}],
\end{equation}
and
\begin{equation}
\bar{E}^\basis[n_{\Psi^\basis},n_{2,\Psi^\basis}] \to \bar{E}^\basis[n_{\Psi^\basis_\text{val}},n_{2,\Psi^\basis_\text{val}}],
\end{equation}
where $\Psi^\basis_\text{val}$ is the wave function $\Psi^\basis$ truncated to the ``valence'' orbital space (i.e., with all core orbitals removed). Accordingly, the local range-separation parameter becomes
\begin{equation}
 \mu^\basis(\br{}) \to \mu^\basis_\text{val}(\br{})
\end{equation}
where
\begin{equation}
 \mu^\basis_\text{val}(\br{}) = \frac{\sqrt{\pi}}{2} W^\basis_\text{val}(\br{}).
\end{equation}
The valence-only effective interaction at coalescence is
\begin{equation}
 \label{eq:wbasis}
 W^\basis_\text{val}(\br{})=
    \begin{cases}
      \frac{\fbasisval}{\twodmrdiagpsival},    & \text{if $\twodmrdiagpsival \ne 0$,}
\\
       \infty,                                                                                          & \text{otherwise,}
    \end{cases}
\end{equation}
where 
\begin{equation}
        \label{eq:fbasisval}
        \fbasisval
        = \sum_{pq}^\basis  \sum_{rstu}^{\basis_\text{val}} w_{pqrs} \Gamma_{rstu}\SO{p}{} \SO{q}{} \SO{t}{} \SO{u}{},
\end{equation}
where $\Gamma_{pqrs} = 2 \elemm{\Psi_\text{loc,val}^\basis}{ \aic{r_\downarrow}\aic{s_\uparrow}\ai{q_\uparrow}\ai{p_\downarrow}}{\Psi_\text{loc,val}^\basis}$ is the opposite-spin two-electron density matrix of the localizing wave function truncated to the valence orbital space $\Psi_\text{loc,val}^{\basis}$, and $\twodmrdiagpsival$ is its associated on-top pair density
\begin{equation}
\label{eq:twordm_val}
 \twodmrdiagpsival = \sum_{pqrs}^{\basis_\text{val}} \Gamma_{pqrs} \SO{p}{} \SO{q}{}  \SO{r}{} \SO{s}{}. 
\end{equation}
In Eqs.~\eqref{eq:fbasisval} and~\eqref{eq:twordm_val}, the indication ``$\basis_\text{val}$'' on top of the sum symbols means that the sums are only over valence (i.e., non-core) orbitals. It is noteworthy that $\mu^\basis_\text{val}(\br{})$ still fulfills Eq.~\eqref{eq:cbs_mu} and thus the frozen-core versions of the basis-set correction functionals still correctly vanish in the CBS limit.

Correspondingly, the frozen-core versions of the effective Hamiltonians in Eqs.~\eqref{eq:def_HeffB} and~\eqref{eq:def_HeffB2} are simply obtained by setting to zero all the one-electron effective integrals $\bar{v}^\basis_{pq}$ and the two-electron effective integrals $\bar{w}^\basis_{pqrs}$ if at least one orbital involved in the integral is a core orbital.

\subsection{Selected configuration interaction to solve the self-consistent eigenvalue equations}
\label{sec:cipsi}

To solve the self-consistent basis-set correction eigenvalue equations [Eqs.~\eqref{eq:HeffB} and \eqref{eq:HeffB2}], we use an adaptation of the configuration-interaction perturbatively selected iteratively (CIPSI) algorithm~\cite{HurMalRan-JCP-73, GinSceCaf-CJC-13, GinSceCaf-JCP-15}, similar to the computational strategy already used in the context of RSDFT\cite{FerGinTou-JCP-18}.

To solve Eq.~\eqref{eq:HeffB} for a given basis set $\basis$, we start at the first iteration, denoted as $k-1$, from a guess wave function $\Psi^{\basis,(k-1)}_0$ (usually a CIPSI ground-state wave function for the standard Hamiltonian)
\begin{equation}
\ket{\Psi^{\basis,(k-1)}_0} = \sum_{{\rm I}\,\in\,\mathcal{R}^{(k-1)}} \,\,c_{\rm I}^{(k-1)} \,\,\ket{\rm I},
\end{equation}
where $\mathcal{R}^{(k-1)}$ denotes a set of Slater determinants. We then use the density of this wave function to form the following effective Hamiltonian at the next iteration $k$ 
\begin{equation}
\hat{H}_\text{eff}^{\basis,(k)}= \hat{H}_\text{eff}^\basis[n_{\Psi_0^{\basis,(k-1)}}],
\end{equation}
and we want to find the associated ground-state wave function $\Psi^{\basis,(k)}$
\begin{equation}
\hat{H}_\text{eff}^{\basis,(k)} \ket{\Psi^{\basis,(k)}} = {\cal E}^{\basis,(k)} \ket{\Psi^{\basis,(k)}}.
\end{equation}
This wave function is obtained by the CIPSI algorithm as
\begin{equation}
\ket{\Psi^{\basis,(k)}} = \sum_{{\rm I}\,\in\,\mathcal{R}^{(k)}} \,\,c_{\rm I} \,\,\ket{\rm I},
\end{equation}
where $\mathcal{R}^{(k)}$ is the new set of Slater determinants at iteration $k$. According the CIPSI algorithm, the set $\mathcal{R}^{(k)}$ is obtained by repeatedly adding to a reference wave function $\Psi^{\basis,(k)}_\text{ref}$ the determinants ${\rm K}$ having the largest second-order perturbation theory (PT2) contributions $|{\cal E}_{{\rm K},\text{PT2}}^{(k)}|$ with
\begin{equation}
\begin{aligned}
  {\cal E}_{{\rm K},\text{PT2}}^{(k)}&= -\frac{\vert\elemm{\Psi^{\basis,(k)}_\text{ref}}{ \hat{H}_\text{eff}^{\basis,(k)} }{{\rm K}}\vert^2}{\elemm{{\rm K}}{\hat{H}_\text{eff}^{\basis,(k)} }{{\rm K}}   - \elemm{\Psi^{\basis,(k)}_\text{ref}}{\hat{H}_\text{eff}^{\basis,(k)} }{\Psi^{\basis,(k)}_\text{ref}}},
\end{aligned}
\end{equation}
iteratively doubling the number of determinants in $\Psi^{\basis,(k)}_\text{ref}$ until the absolute value of the total PT2 energy correction due to the missing determinants
\begin{equation}
\begin{aligned}
|{\cal E}_{\text{PT2}}^{(k)}|&= \left|\sum_{{{\rm K}}\,\not\in\,\mathcal{R}^{(k)}} {\cal E}_{{\rm K},\text{PT2}}^{(k)} \right|,
\end{aligned}
\end{equation}
is smaller than a given threshold. To reduce the cost of the evaluation of the PT2 contribution, the semi-stochastic multi-reference approach of Garniron \textit{et al.} \cite{GarSceLooCaf-JCP-17} is used, adopting the technical specifications recommended in that work. 

This determines the set of determinants $\mathcal{R}^{(k)}$ which is then fixed for the rest of the iteration $k$. The energy $E_0^{\basis,(k)}$ for this iteration is then determined according to the minimization in Eq.~\eqref{eq:E_minpsi}
\begin{equation}
  \begin{aligned}
 E_0^{\basis,(k)} = \min_{\Psi^{\basis,(k)}} \Big\{\elemm{\Psi^{\basis,(k)}}{\kinop + \weeop +  \hat{V}_\text{ne}}{\Psi^{\basis,(k)}} \\+ \bar{E}^\basis[n_{\Psi^{\basis,(k)}}]\Big\},
  \end{aligned}
\end{equation}
which amounts to solving the iterative equation
\begin{equation}
\hat{H}_\text{eff}^{\basis}[n_{\Psi_0^{\basis,(k)}}]  \ket{\Psi_0^{\basis,(k)}} = {\cal E}_0^{\basis,(k)} \ket{\Psi_0^{\basis,(k)}} ,
\end{equation}
for the optimal coefficients of the determinants $\{c_{\rm I}^{(k)} \}$ leading to the minimizing wave function at iteration $k$
\begin{equation}
\ket{\Psi^{\basis,(k)}_0} = \sum_{{\rm I}\,\in\,\mathcal{R}^{(k)}} \,\,c_{\rm I}^{(k)} \,\,\ket{\rm I}.
\end{equation}
The iterations over $k$ are repeated until the variation of $E_0^{\basis,(k)}$ is smaller than a given threshold. The evaluation of the dipole moment is obtained as the expectation value of the dipole operator over the converged wave function $\Psi^\basis_0$. 

The same approach is used for solving Eq.~\eqref{eq:HeffB2} which involves the on-top pair density in addition to the density.

%%%%%%%%%%%%%%%%%%%%%%%%%%%%%%%%%%%%%%%%%%%%%%%%%%%%%%%%%%%%%%%%%%%%%%%%%%%
\section{Computation of total energies and dipole moments}
\label{sec:numerical}

\subsection{Computational details}

We study the total ground-state energies of the Be atom and BH molecule together with the dipole moments of the BH, FH, H$_2$O molecules in their ground states and of the CH$_2$ molecule in its lowest spin-singlet state. We report standard CIPSI (i.e., near FCI) results without the basis-set correction (referred to as ``CIPSI''), as well as CIPSI results including the basis-set correction using the PBE-UEG and PBE-OT functionals with or without self-consistency. The non-self-consistent calculations are referred to as ``CISPI+PBE-UEG and ``CISPI+PBE-OT'', whereas the self-consistent calculations are referred to as ``SC CISPI+PBE-UEG'' and ``SC CISPI+PBE-OT'' where SC stands for self-consistent. The orbitals used for all converged CIPSI calculations are the natural orbitals obtained from a standard CISPI calculation stopped at a total PT2 energy correction smaller in absolute value than $0.001$ Ha. For the localizing wave function $\Psi_\text{loc}^\basis$ involved in the definition of the local range-separation parameter $\mu^\basis(\br{})$ (see Section \ref{sec:mur}), we choose either a single Slater determinant (SD) built from the natural orbitals of the largest CIPSI wave function (which we refer as $\murhf$) or the largest CIPSI wave function (which we refer as $\murfci$). We use the Dunning correlation-consistent basis-set family~\cite{Dun-JCP-89,WooDun-JCP-93,WooDun-JCP-95,PetDun-JCP-02,PriAltDidGibWin-JCIM-19}. We perform both non-frozen-core calculations using the core-valence aug-cc-pCV$X$Z basis sets, and frozen-core calculations (with the 1s orbitals of non-hydrogen atoms frozen) using the valence aug-cc-pV$X$Z basis sets and the corresponding frozen-core version of the basis-set correction (see Section \ref{sec:FC}). All the CIPSI calculations have been performed with \textsc{Quantum Package}\cite{QP2}. 

We also report the dipole moment at the coupled cluster singles doubles perturbative triples [CCSD(T)] level which were taken from Ref.~\onlinecite{HalKloHelJor-JCP-99} for the BH and FH molecules, and obtained using linear-response calculations from the \textsc{Dalton} software\cite{dalton2013,Dal-PROG-20} for the CH$_2$ and H$_2$O molecules. The molecular geometries are taken from Ref.~\onlinecite{HalKloHelJor-JCP-99} for BH and FH, and from Ref.~\onlinecite{HaiHea-JCTC-18} for H$_2$O and CH$_2$.

%%%%%%%%%%%%%%%%%%%%%%%%%%%%%%%%%%%%%%%%%%%%%%%%%%%%%%%%%%%%%%%%%%%%%%%%%%%%%%%%%%%%%%%%%%%%%%%%%%%%%%%%%%%%%%%%%%%%
\begin{table*}
\caption{Total ground-state energies (in Ha) of the Be atom calculated using the aug-cc-pCV$X$Z (ACV$X$Z) basis sets (with $X$ = D, T, Q) with CIPSI without the frozen-core approximation and including different basis-set corrections with or without self-consistency. The energy lowering $\Delta E_{\text{SC}}$ (in ${\mu}$Ha) from the non-self-consistent to the self-consistent version of the basis-set correction is reported in square brackets.}
 \label{atom-tot}
\begin{ruledtabular}
 \begin{tabular}{llll}
                                    &      \multicolumn{1}{c}{   ACVDZ}   &         \multicolumn{1}{c}{ACVTZ}     &  \multicolumn{1}{c}{ACVQZ}           \\
 \hline
  CIPSI                                 & -14.6519225 & -14.6623971 & -14.6655767   \\[0.1cm]
 $\levelcomp{PBE-UEG}{\murhf}$              & -14.6683617  & -14.6686314 & -14.6681020   \\
 SC $\levelcomp{PBE-UEG}{\murhf}$ [$\Delta E_{\text{SC}}$]  & -14.6683878 [-26.1]        & -14.6686354 [-4.0]        & -14.6681026 [-0.6]          \\[0.1cm]
 $\levelcomp{PBE-UEG}{\murfci}$             & -14.6677035  & -14.6683762 & -14.6679643   \\
 SC $\levelcomp{PBE-UEG}{\murfci}$ [$\Delta E_{\text{SC}}$]  & -14.6677395 [-36.0]        & -14.6683874 [-11.2]  & -14.6679681 [-3.8]          \\[0.1cm]
 $\levelcomp{PBE-OT}{\murhf}$               & -14.6663376  & -14.6678846 & -14.6677982   \\
 SC $\levelcomp{PBE-OT}{\murhf}$ [$\Delta E_{\text{SC}}$]  & -14.6663741 [-36.5]     & -14.6678956 [-11.0]       & -14.6678020 [-3.8]          \\[0.1cm]
 $\levelcomp{PBE-OT}{\murfci}$              & -14.6659463  & -14.6677128& -14.6677140  \\
 SC $\levelcomp{PBE-OT}{\murfci}$ [$\Delta E_{\text{SC}}$]  & -14.6659748 [-28.5]     & -14.6677223 [-9.5]       & -14.6677171 [-3.1]          \\
\hline
 \multicolumn{4}{c}{Exact non-relativistic total energy$^a$} \\   
 \multicolumn{4}{c}{-14.6673565}
\end{tabular}
\end{ruledtabular}
{\raggedright  $^a$From Ref.~\onlinecite{HorAdaBub-PRA-19}. \par}
 \end{table*}
%%%%%%%%%%%%%%%%%%%%%%%%%%%%%%%%%%%%%%%%%%%%%%%%%%%%%%%%%%%%%%%%%%%%%%%%%%%%%%%%%%%%%%%%%%%%%%%%%%%%%%%%%%%%%%%%%%%%

%%%%%%%%%%%%%%%%%%%%%%%%%%%%%%%%%%%%%%%%%%%%%%%%%%%%%%%%%%%%%%%%%%%%%%%%%%%%%%%%%%%%%%%%%%%%%%%%%%%%%%%%%%%%%%%%%%%%
 \begin{table*}
\caption{Total ground-state energies (in Ha) of the BH molecule calculated using the aug-cc-pCV$X$Z (ACV$X$Z) basis sets (with $X$ = D, T, Q, 5) with CIPSI without the frozen-core approximation and including different basis-set corrections with or without self-consistency. The energy lowering $\Delta E_{\text{SC}}$ (in ${\mu}$Ha) from the non-self-consistent to the self-consistent version of the basis-set correction is reported in square brackets.}
 \label{bh-tot}
\begin{ruledtabular}
 \begin{tabular}{lllll}
                                    &      \multicolumn{1}{c}{   ACVDZ}   &         \multicolumn{1}{c}{ACVTZ}     &  \multicolumn{1}{c}{ACVQZ}     & \multicolumn{1}{c}{ACV5Z}              \\
 \hline
  CIPSI                                 & -25.2550150 & -25.2786179   &  -25.2857583  &    -25.2873779   \\[0.1cm]
 $\levelcomp{PBE-UEG}{\murhf}$              & -25.2838179 & -25.2896303   &  -25.2906772  &    -25.2900079   \\
 SC $\levelcomp{PBE-UEG}{\murhf}$ [$\Delta E_{\text{SC}}$]  & -25.2839270 [-109.1]       & -25.2896471 [-16.8]         &  -25.2906804 [-3.2]          &           --       \\[0.1cm]
 $\levelcomp{PBE-UEG}{\murfci}$             & -25.2825079 & -25.2890245   &  -25.2903975  &           --       \\
 SC $\levelcomp{PBE-UEG}{\murfci}$ [$\Delta E_{\text{SC}}$]  & -25.2826090 [-101.1]       & -25.2890400 [-15.5]         &     --          &           --       \\[0.1cm]
 $\levelcomp{PBE-OT}{\murhf}$               & -25.2798297 & -25.2880486   &  -25.2899774  &           --       \\
 SC $\levelcomp{PBE-OT}{\murhf}$ [$\Delta E_{\text{SC}}$]  & -25.2800391 [-209.4]       & -25.2881008 [-52.2]         &  -25.2899937 [-16.3]         &           --       \\[0.1cm]
 $\levelcomp{PBE-OT}{\murfci}$              & -25.2789738 & -25.2876363   &  -25.2897883  &           --       \\
 SC $\levelcomp{PBE-OT}{\murfci}$ [$\Delta E_{\text{SC}}$]  & -25.2791600 [-186.2]       & -25.2876809 [-44.6]         &     --          &           --       \\
\hline
                                        \multicolumn{5}{c}{CIPSI total energy extrapolated to the CBS limit} \\   
                                        \multicolumn{5}{c}{ -25.289032        } 
\end{tabular}
\end{ruledtabular}
 \end{table*}
%%%%%%%%%%%%%%%%%%%%%%%%%%%%%%%%%%%%%%%%%%%%%%%%%%%%%%%%%%%%%%%%%%%%%%%%%%%%%%%%%%%%%%%%%%%%%%%%%%%%%%%%%%%%%%%%%%%%%%%%%%%%%%%%%%

\subsection{Total energies of the Be atom and the BH molecule}
\label{tot_energies}

Tables \ref{atom-tot} and \ref{bh-tot} report the total energies of the Be atom and the BH molecule, respectively, calculated using the aug-cc-pCV$X$Z basis sets with CIPSI without any basis-set correction and with different basis-set corrections. It can be observed that the total energies obtained with the basis-set corrections converge much faster toward the estimated exact total energies than the total energies obtained without any basis-set correction. For the Be atom, all the basis-set corrected total energies from the aug-cc-pCVDZ to the aug-cc-pCVQZ basis sets have errors below 1.6 mHa $\approx$ 1 kcal/mol compared to the estimated exact total energy, whereas without basis-set correction such an accuracy is barely reached only with the aug-cc-pCVQZ basis set. Similar trends are observed for the BH molecule: all the basis-set corrected total energies have errors below 1 kcal/mol already from the aug-cc-pCVTZ basis set, whereas without basis-set correction such an accuracy is barely reached with the aug-cc-pCV5Z basis set.

Focusing now on the differences between the various basis-set corrections, we can notice that i) using the local range-separation parameter $\murhf$ gives larger basis-set corrections than using $\murfci$, ii) the PBE-UEG functional gives larger basis-set corrections than the PBE-OT functional. As regards the effect of the self-consistency, it is remarkable to notice that self-consistency lowers the total energy by a very small fraction of the total basis-set correction (typically less than $1\%$), whatever the choice of functional or local range-separation parameter. These results thus validate the previously introduced non-self-consistent approximation to the basis-set correction (see Section~\ref{sec:nonself}) for energy calculations.

%%%%%%%%%%%%%%%%%%%%%%%%%%%%%%%%%%%%%%%%%%%%%%%%%%%%%%%%%%%%%%%%%%%%%%%%%%%%%%%%%%%%%%%%%%%%%%%%%%%%%%%%%%%%%%%%%%%%
\begin{table*}
\caption{Dipole moment (in a.u.) of the ground state of the BH molecule calculated using the aug-cc-pCV$X$Z (ACV$X$Z) basis sets (with $X$ = D, T, Q, 5) by Hartree-Fock (HF), CCSD(T), and CIPSI without the frozen-core approximation and including different self-consistent basis-set corrections. For the CIPSI calculations, the PT2 energy correction $|{\cal E}_\text{PT2}|$ (in Ha) is reported in square brackets. Extrapolations to the CBS limit are given in the last column.}
\label{bh_acvx_cc}
\begin{ruledtabular}
\begin{tabular}{llllll}
                            &      \multicolumn{1}{c}{   ACVDZ}   &         \multicolumn{1}{c}{ACVTZ}     &  \multicolumn{1}{c}{ACVQZ}     & \multicolumn{1}{c}{ACV5Z} & \multicolumn{1}{c}{CBS}             \\
\hline                                                                  
HF                          &    0.68796                          &          0.68570                      &     0.68489                        &    0.68530                 &   \\
CCSD(T)$^a$ &    0.52968                          &          0.54649                      &     0.54984                        &    0.55125                 & 0.55271 \\
CIPSI $\scriptstyle[|{\cal E}_\text{PT2}|]$    &    0.52758 $\PT{2 \times 10^{-6}}$     &          0.54388   $\PT{4 \times 10^{-6}}$ &     0.54789   $\PT{4 \times 10^{-6}}$ &    0.54975   $\PT{2 \times 10^{-5}}$ & 0.55142 \\
CIPSI extrapolated to ${\cal E}_\text{PT2}\to 0$                             & 0.52757       &          0.54386                      &     0.54790                         &    0.54967                &   0.55126\\[0.3cm]
SC $\levelcomp{PBE-UEG}{\murhf}$ $\scriptstyle[|{\cal E}_\text{PT2}|]$   &  0.53658 $\PT{2 \times 10^{-6}}$ &  0.54835 $\PT{4 \times 10^{-6}}$ & 0.55040 $\PT{4 \times 10^{-5}}$ \\
SC $\levelcomp{PBE-UEG}{\murfci}$ $\scriptstyle[|{\cal E}_\text{PT2}|]$  &  0.53758 $\PT{2 \times 10^{-6}}$ & 0.54812 $\PT{4 \times 10^{-6}}$  \\
SC $\levelcomp{PBE-OT}{\murhf}$ $\scriptstyle[|{\cal E}_\text{PT2}|]$    &  0.54340 $\PT{2 \times 10^{-6}}$ & 0.55092 $\PT{4 \times 10^{-6}}$ & 0.55093 $\PT{1 \times 10^{-5}}$ \\
SC $\levelcomp{PBE-OT}{\murfci}$ $\scriptstyle[|{\cal E}_\text{PT2}|]$   &  0.54333 $\PT{2 \times 10^{-6}}$ & 0.54973 $\PT{8 \times 10^{-6}}$
\end{tabular}
\end{ruledtabular}
{\raggedright  $^a$From Ref.~\onlinecite{HalKloHelJor-JCP-99}. \par}
\end{table*}

%%%%%%%%%%%%%%%%%%%%%%%%%%%%%%%%%%%%%%%%%%%%%%%%%%%%%%%%%%%%%%
 \begin{table*}
 \caption{Dipole moment (in a.u.) of the ground state of the BH molecule calculated using the aug-cc-pV$X$Z (AV$X$Z) basis sets (with $X$ = D, T, Q, 5) by Hartree-Fock (HF), CCSD(T), and CIPSI with the frozen-core approximation and including different self-consistent basis-set corrections. For the CIPSI calculations, the PT2 energy correction $|{\cal E}_\text{PT2}|$ (in Ha) is reported in square brackets. Extrapolations to the CBS limit are given in the last column.}
 \label{table_bh_fc_acvxz}
\begin{ruledtabular}
 \begin{tabular}{llllll}
                  &      \multicolumn{1}{c}{   AVDZ}   &         \multicolumn{1}{c}{AVTZ}     &  \multicolumn{1}{c}{AVQZ}     & \multicolumn{1}{c}{AV5Z} & \multicolumn{1}{c}{CBS}             \\
\hline
HF                & 0.68796 & 0.68650 & 0.68494 & 0.68496 \\
CCSD(T)           & 0.52939 & 0.54500 & 0.54724 & 0.54843 & 0.54966\\
CIPSI $\scriptstyle[|{\cal E}_\text{PT2}|]$    & 0.52782 $\PT{3 \times 10^{-8}}$ & 0.54334 $\PT{1 \times 10^{-7}}$ & 0.54563 $\PT{4 \times 10^{-7}}$ & 0.54691 $\PT{1 \times 10^{-6}}$ & 0.54823 \\[0.2cm]
SC $\levelcomp{PBE-UEG}{\murhf}$ $\scriptstyle[|{\cal E}_\text{PT2}|]$ & 0.53791 $\PT{4 \times 10^{-7}}$ & 0.54815 $\PT{7 \times 10^{-7}}$ & 0.54790 $\PT{3 \times 10^{-6}}$ & 0.54815 $\PT{3 \times 10^{-6}}$ \\
SC $\levelcomp{PBE-OT}{\murhf}$ $\scriptstyle[|{\cal E}_\text{PT2}|]$    & 0.54512 $\PT{4 \times 10^{-7}}$ & 0.55029 $\PT{5 \times 10^{-7}}$ & 0.54880 $\PT{2 \times 10^{-8}}$ \\
 \end{tabular}
\end{ruledtabular}
 \end{table*}

%%%%%%%%%%%%%%%%%%%%%%%%%%%%%%%%%%%%%%%%%%%%%%%%%%%%%%%%%%%%%%
\begin{figure*}
\includegraphics[width=0.45\linewidth]{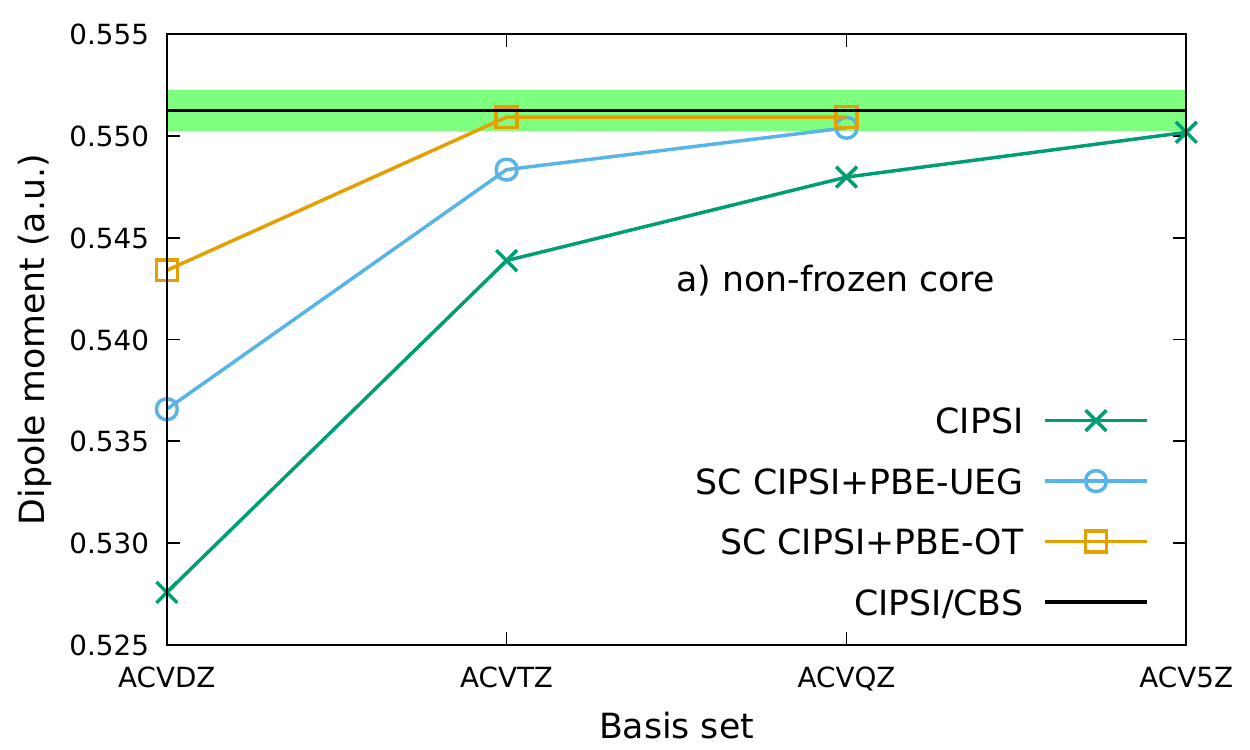}
\includegraphics[width=0.45\linewidth]{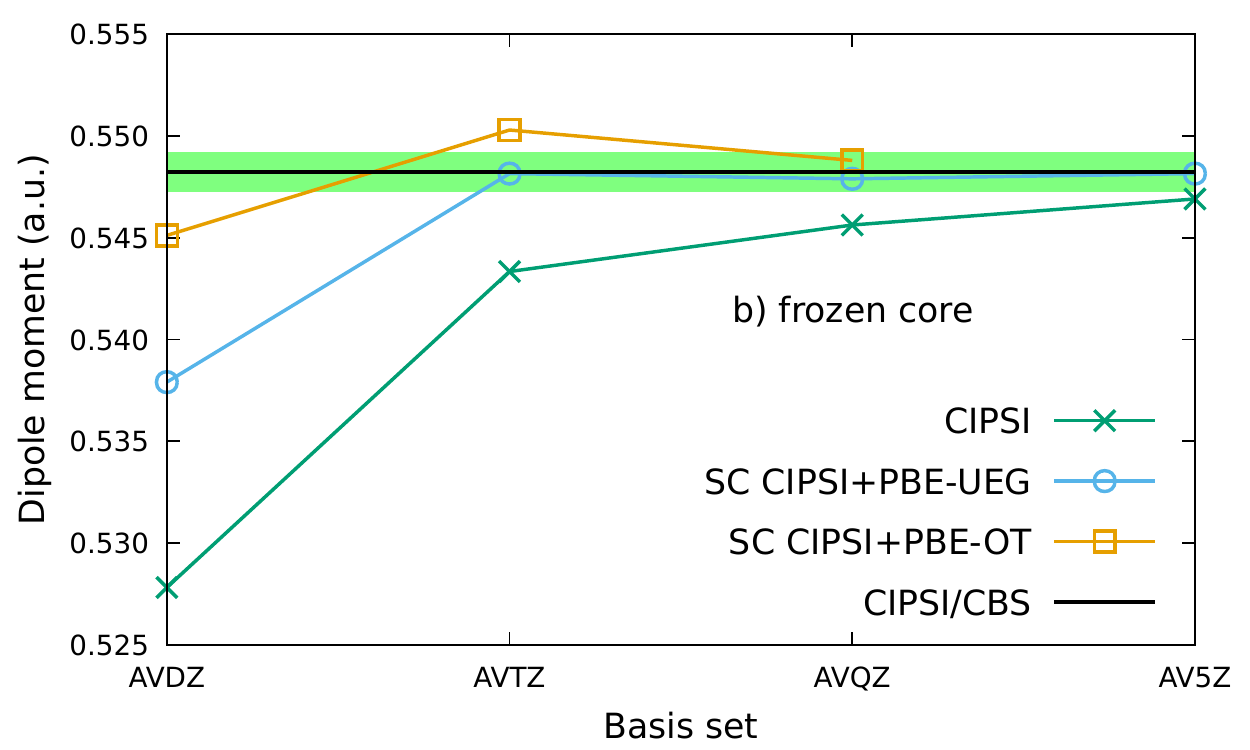}
\caption{Basis-set convergence of the dipole moment of the ground state of the BH molecule calculated using the aug-cc-pCV$X$Z (ACV$X$Z) and aug-cc-pV$X$Z (AV$X$Z) basis sets (with $X$ = D, T, Q, 5) by CIPSI including different self-consistent basis-set corrections without the frozen-core approximation (a) and with the frozen-core approximation (b). The self-consistent basis-set corrections are for the local range-separation parameter $\mu_{\text{SD}}$. The green area indicates an error of $\pm 0.001$ a.u. around the CIPSI CBS value.}
 \label{plot_bh_bc_acvxz}
\end{figure*}
%%%%%%%%%%%%%%%%%%%%%%%%%%%%%%%%%%%%%%%%%%%%%%%%%%%%%%%%%%%%%%

\subsection{Dipole moments of the BH, FH, H$_2$O, and CH$_2$ molecules}
\label{sec:BH}
As seen from Section \ref{tot_energies}, the self-consistency of the basis-set correction does not lead to significant changes of the total energies. Nevertheless, one can wonder if the effective wave functions obtained with the self-consistent basis-set correction provide better properties. 

We choose to investigate this aspect through the computation of the dipole moments of the BH, FH, H$_2$O, and CH$_2$ molecules for several reasons: i) the basis-set convergence of dipole moments with correlated wave-function methods is known to be slow (see, e.g., Refs.~\onlinecite{HalKloHelJor-JCP-99,BakGauHelJorOls-CPL-00,HaiHea-JCTC-18}), ii) these molecules are sufficiently small to have near-CBS reference values, iii) the BH and CH$_2$ molecules exhibit a non-trivial mixture of both strong and weak correlation due to the 2s-2p near degeneracy in the boron and carbon atoms, whereas the FH and H$_2$O molecules are dominated by weak correlation effects. 

\subsubsection{Dipole moment of the BH molecule}
\label{calc_all_elec}

We start by a detailed analysis on the BH molecule. The slow convergence of the dipole moment of the BH molecule with respect to the size of the basis set has been illustrated at various correlation levels including CCSD(T) by Halkier \textit{et al.}~\cite{HalKloHelJor-JCP-99}, and we report in Table \ref{bh_acvx_cc} their CCSD(T) results at the non-frozen core level. In order to have a reasonable estimate of the dipole moment in the CBS limit, we use the two-point $X^{-3}$ extrapolation proposed in Ref.~\onlinecite{HalKloHelJor-JCP-99} using the aug-cc-pCVQZ and aug-cc-pCV5Z basis sets. At the CCSD(T) level, an error of about 0.001 a.u. with respect to the CBS limit is barely reached with the aug-cc-pCV5Z basis set, while the error is about 0.006 a.u. and 0.003 a.u. with the aug-cc-pCVTZ and aug-cc-pCVQZ basis set, respectively, showing indeed the quite slow convergence of the dipole moment of BH.

We also report in Table \ref{bh_acvx_cc} the dipole moment at the CIPSI level together with the value of the PT2 energy correction $|{\cal E}_\text{PT2}|$ associated with the variational wave function for which the dipole moment have been calculated. The values of $|{\cal E}_\text{PT2}|$ are all below $2 \times 10^{-5}$ Ha, which was found to be mandatory to converge the CIPSI dipole moments to a precision of about 0.0001 a.u. for the BH molecule. These represent therefore rather large calculations involving about $10^8$ Slater determinants for the largest aug-cc-pCV5Z basis set. In order to obtain an estimation of the error of the dipole moment at the CIPSI level in a given basis set with respect to FCI, we extrapolate the CIPSI dipole moment to $|{\cal E}_\text{PT2}|\to 0$ using a linear extrapolation as a function of $|{\cal E}_\text{PT2}|$ (similar to the proposal of Holmes \textit{et al.}~\cite{HolUmrSha-JCP-17} for the total energy) using different values of $|{\cal E}_\text{PT2}|$. As one can see from Table \ref{bh_acvx_cc}, for any basis set, the difference between the dipole moment calculated by CIPSI with the smallest $|{\cal E}_\text{PT2}|$ available and the dipole moment extrapolated with respect to $|{\cal E}_\text{PT2}|$ are negligible. Thus, with the values of $|{\cal E}_\text{PT2}|$ used, the CIPSI dipole moments provide reliable estimates of the FCI dipole moments. As regards basis-set errors, similar to the CCSD(T) results, an error of about 0.001 a.u. on the CIPSI dipole moment with respect to the CBS limit is barely reached only using the aug-cc-pCV5Z basis set, illustrating once more the slow convergence of the dipole moment with respect to the size of the basis set.

From Table \ref{bh_acvx_cc} one can also notice that there is a small deviation between the CCSD(T) and the CIPSI dipole moments which is of about 0.002 a.u. for the aug-cc-pCV$X$Z basis sets with $X$= D, T, Q and of about 0.001 a.u. for the aug-cc-pCV5Z basis set. Also, the deviation between the CCSD(T) and CIPSI dipole moments extrapolated to the CBS limit is about 0.001 a.u., showing that CCSD(T) provides a very accurate value for the dipole moment of the BH molecule. 

Coming now to the self-consistent basis-set correction calculations, we report in Table \ref{bh_acvx_cc} the dipole moments obtained using the $\pbeuegXi$ and $\pbeontns$ functionals using the local range-separation parameters $\murhf$ and $\murfci$, and a graphical representation of the data for $\murhf$ is given in Figure~\ref{plot_bh_bc_acvxz} (panel a). Clearly, the basis-set correction strongly accelerates the convergence to the CBS limit. More specifically, it can be observed that i) in a given basis set, all basis-set corrections significantly improve the dipole moment with respect to the CIPSI value, ii) the $\pbeontns$ functional gives more accurate results than the $\pbeuegXi$ functional, iii) an error of about 0.001 a.u. on the dipole moment is obtained already with the aug-cc-pCVTZ basis set when using the $\pbeontns$ functional. The result i) shows that, although the self-consistency does not lead to significant improvement on the total energy (see Table \ref{bh-tot}), it is crucial to yield effective wave functions providing better properties, illustrating the impact and the accuracy of the effective basis-set correction potentials employed. The result ii) shows that the use of the on-top pair density of the wave function rather than that estimated from the UEG gives a better approximation of the exact on-top pair density, which can be understood as a signature of strong-correlation effects.

Finally, we also report in Table \ref{table_bh_fc_acvxz} the dipole moments obtained with the frozen-core approximation using the aug-cc-pV$X$Z basis sets, and the corresponding graphical representation is given in Figure~\ref{plot_bh_bc_acvxz} (panel b). Again, the basis-set correction with either the $\pbeuegXi$ or $\pbeontns$ functional yields a faster basis-set convergence of the dipole moment that in the standard CIPSI calculations. However, in this case, the convergence toward the CBS limit is slightly better when using the $\pbeuegXi$ functional. The $\pbeontns$ functional slightly overestimates the dipole moment by $0.002$ a.u. and $0.0006$ a.u. with the aug-cc-pVTZ and aug-cc-pVQZ basis sets, respectively, whereas the $\pbeuegXi$ functional yields a deviation below $0.0005$ a.u. from the aug-cc-pVTZ to the aug-cc-pV5Z basis set.

%%%%%%%%%%%%%%%%%%%%%%%%%%%%%%%%%%%%%%%%%%%%%%%%%%%%%%%%%%%%%%%%%%%%%%%%%%%%%%%%%%%%%%%%%%%%%%%%%%%%%%%%%%%%%%%%%%%%%%%%%%%
\subsection{Dipole moments of the FH, H$_2$O, and CH$_2$ molecules}
\label{sec:FH}

We now pursue our analysis on the FH, H$_2$O, and CH$_2$ molecules using only the frozen-core approximation. The basis-set convergence of the dipole moments of these molecules was studied in Refs.~\onlinecite{HalKloHelJor-JCP-99,BakGauHelJorOls-CPL-00} at the CCSD(T) level. In Tables \ref{fh_avx_cc}, \ref{h2o_avx_cc}, and \ref{ch2_avx_cc}, we report CCSD(T) and CIPSI results with the aug-cc-pV$X$Z basis sets. The CBS estimates are obtained from a two-point $X^{-3}$ extrapolation using the aug-cc-pVQZ and aug-cc-pV5Z basis sets at the CCSD(T) level, and also at the CIPSI level in the case of CH$_2$. One can notice that, at the CCSD(T) or CIPSI level, an error of about 0.001 a.u. with respect to the estimated CBS limit is barely reached only with the aug-cc-pV5Z basis set. Also, in the case of CH$_2$, there is a significant discrepancy between the extrapolated CCSD(T) and CIPSI dipole moments, which might be due to some strong-correlation effects that are mistreated at the CCSD(T) level.

%%%%%%%%%%%%%%%%%%%%%%%%%%%%%%%%%%%%%%%%%%%%%%%%%%%%%%%%%%%%%%%%%%%%%%%%%%%%%%%%%%%%%%%%%%%%%%%
 \begin{table*}
 \caption{Dipole moment (in a.u.) of the ground state of the FH molecule calculated using the aug-cc-pV$X$Z (AV$X$Z) basis sets (with $X$ = D, T, Q, 5) by Hartree-Fock (HF), CCSD(T), and CIPSI with the frozen-core approximation and including different self-consistent basis-set corrections. For the CIPSI calculations, the PT2 energy correction $|{\cal E}_\text{PT2}|$ (in Ha) is reported in square brackets. Extrapolations to the CBS limit are given in the last column.}
 \label{fh_avx_cc}
\begin{ruledtabular}
 \begin{tabular}{llllll}
                  &      \multicolumn{1}{c}{   AVDZ}   &         \multicolumn{1}{c}{AVTZ}     &  \multicolumn{1}{c}{AVQZ}     & \multicolumn{1}{c}{AV5Z} & \multicolumn{1}{c}{CBS}             \\
\hline
HF                & 0.75976 & 0.75750 & 0.75634 & 0.75617\\
CCSD(T)           & 0.70342 & 0.70465 & 0.70707 & 0.70794 & 0.70903\\
CIPSI $\scriptstyle[|{\cal E}_\text{PT2}|]$    & 0.70249 $\PT{9 \times 10^{-6}}$ & 0.70406 $\PT{1 \times 10^{-4}}$ & 0.70662 $\PT{1 \times 10^{-4}}$ \\
CIPSI extrapolated wrt ${\cal E}_\text{PT2}$   & 0.70248 & 0.70391 & 0.70646 \\[0.1cm]

SC $\levelcomp{PBE-UEG}{\murhf}$ $\scriptstyle[|{\cal E}_\text{PT2}|]$ & 0.71326 $\PT{3 \times 10^{-5}}$ & 0.70873 $\PT{1 \times 10^{-4}}$ \\
SC $\levelcomp{PBE-OT}{\murhf}$ $\scriptstyle[|{\cal E}_\text{PT2}|]$  & 0.71362 $\PT{2 \times 10^{-5}}$ & 0.70915 $\PT{1 \times 10^{-4}}$
 \end{tabular}
\end{ruledtabular}
 \end{table*}
%%%%%%%%%%%%%%%%%%%%%%%%%%%%%%%%%%%%%%%%%%%%%%%%%%%%%%%%%%%%%%%%%%%%%%%%%%%%%%%%%%%%%%%%%%%%%%%

%%%%%%%%%%%%%%%%%%%%%%%%%%%%%%%%%%%%%%%%%%%%%%%%%%%%%%%%%%%%%%%%%%%%%%%%%%%%%%%%%%%%%%%%%%%%%%%
 \begin{table*}
 \caption{Dipole moment (in a.u.) of the ground state of the H$_2$O molecule calculated using the aug-cc-pV$X$Z (AV$X$Z) basis sets (with $X$ = D, T, Q, 5) by Hartree-Fock (HF), CCSD(T), and CIPSI with the frozen-core approximation and including different self-consistent basis-set corrections. For the CIPSI calculations, the PT2 energy correction $|{\cal E}_\text{PT2}|$ (in Ha) is reported in square brackets. Extrapolations to the CBS limit are given in the last column.}
 \label{h2o_avx_cc}
\begin{ruledtabular}
 \begin{tabular}{llllll}
                  &      \multicolumn{1}{c}{   AVDZ}   &         \multicolumn{1}{c}{AVTZ}     &  \multicolumn{1}{c}{AVQZ}     & \multicolumn{1}{c}{AV5Z} & \multicolumn{1}{c}{CBS}             \\
\hline
HF                & 0.78670 & 0.78038 & 0.77955 & 0.77956\\
CCSD(T)           & 0.72703 & 0.72364 & 0.72695 & 0.72815 & 0.72941\\
CIPSI $\scriptstyle[|{\cal E}_\text{PT2}|]$    & 0.72610 $\PT{3 \times 10^{-5}}$ & 0.72294 $\PT{2 \times 10^{-4}}$ \\[0.1cm]

SC $\levelcomp{PBE-UEG}{\murhf}$ $\scriptstyle[|{\cal E}_\text{PT2}|]$ & 0.73809 $\PT{2 \times 10^{-5}}$ &  0.72818   $\PT{2 \times 10^{-5}}$      \\
SC $\levelcomp{PBE-OT}{\murhf}$ $\scriptstyle[|{\cal E}_\text{PT2}|]$  & 0.73840 $\PT{2 \times 10^{-4}}$ &  0.72855 $\PT{1 \times 10^{-4}}$      \\
 \end{tabular}
\end{ruledtabular}
 \end{table*}
%%%%%%%%%%%%%%%%%%%%%%%%%%%%%%%%%%%%%%%%%%%%%%%%%%%%%%%%%%%%%%%%%%%%%%%%%%%%%%%%%%%%%%%%%%%%%%%

%%%%%%%%%%%%%%%%%%%%%%%%%%%%%%%%%%%%%%%%%%%%%%%%%%%%%%%%%%%%%%%%%%%%%%%%%%%%%%%%%%%%%%%%%%%%%%%
 \begin{table*}
 \caption{Dipole moment (in a.u.) of the lowest spin-singlet state of the CH$_2$ molecule calculated using the aug-cc-pV$X$Z (AV$X$Z) basis sets (with $X$ = D, T, Q, 5) by Hartree-Fock (HF), CCSD(T), and CIPSI with the frozen-core approximation and including different self-consistent basis-set corrections. For the CIPSI calculations, the PT2 energy correction $|{\cal E}_\text{PT2}|$ (in Ha) is reported in square brackets. Extrapolations to the CBS limit are given in the last column.}
 \label{ch2_avx_cc}
\begin{ruledtabular}
 \begin{tabular}{llllll}
                  &      \multicolumn{1}{c}{   AVDZ}   &         \multicolumn{1}{c}{AVTZ}     &  \multicolumn{1}{c}{AVQZ}     & \multicolumn{1}{c}{AV5Z} & \multicolumn{1}{c}{CBS}             \\
\hline
HF                & 0.74878 & 0.74478 & 0.74355 & 0.74353\\
CCSD(T)           & 0.65614 & 0.66009 & 0.66211 & 0.66310 & 0.66416\\
CIPSI $\scriptstyle[|{\cal E}_\text{PT2}|]$    & 0.65120   $\PT{2 \times 10^{-5}}$ & 0.65446   $\PT{3 \times 10^{-5}}$ & 0.65643   $\PT{4 \times 10^{-5}}$ & 0.65780   $\PT{1 \times 10^{-4}}$ & 0.65926\\[0.1cm]

SC $\levelcomp{PBE-UEG}{\murhf}$ $\scriptstyle[|{\cal E}_\text{PT2}|]$ & 0.66249 $\PT{2 \times 10^{-5}}$ & 0.65958 $\PT{3 \times 10^{-5}}$ & 0.65890 $\PT{3 \times 10^{-5}}$\\
SC $\levelcomp{PBE-OT}{\murhf}$ $\scriptstyle[|{\cal E}_\text{PT2}|]$  & 0.66527 $\PT{2 \times 10^{-5}}$ & 0.66055 $\PT{4 \times 10^{-5}}$ & 0.65932 $\PT{5 \times 10^{-4}}$
 \end{tabular}
\end{ruledtabular}
 \end{table*}
%%%%%%%%%%%%%%%%%%%%%%%%%%%%%%%%%%%%%%%%%%%%%%%%%%%%%%%%%%%%%%%%%%%%%%%%%%%%%%%%%%%%%%%%%%%%%%%

We also report in Tables \ref{fh_avx_cc}, \ref{h2o_avx_cc}, and \ref{ch2_avx_cc}, the results using the self-consistent basis-set correction. In contrast with the BH molecule, the effective wave functions obtained with the $\pbeuegXi$ and $\pbeontns$ functionals yield always very similar dipole moments for the FH and H$_2$O molecules. This can be explained by the fact that these molecules at their equilibrium geometries are weakly correlated systems for which the on-top pair density based on the UEG is a good approximation. For the FH, H$_2$O, and CH$_2$ molecules, the dipole moment is overestimated with the aug-cc-pVDZ basis set using both functionals, but the results with the aug-cc-pVTZ basis set are already very close to the estimated CBS limit. From a quantitative point of view, for the FH molecule with the aug-cc-pVDZ basis set the error with respect to the CBS dipole moment is reduced from about 0.007 a.u. at the CIPSI level to about 0.004 a.u. with the basis-set correction, whereas with the aug-cc-pVTZ basis set the error is reduced from 0.005 a.u. to below 0.0005 a.u.. For the CH$_2$ molecule, with the aug-cc-pVDZ basis set the error with respect to the CBS extrapolated CIPSI value is about 0.008 a.u. whereas it is about 0.003 a.u. and 0.006 a.u. using the $\pbeuegXi$ and $\pbeontns$ basis-set corrections, respectively. With the aug-cc-pVTZ basis set, the error at the CIPSI level is still of about 0.005 a.u., whereas it is below 0.001 a.u. for both the $\pbeuegXi$ and $\pbeontns$ functionals. Finally, in the case of the H$_2$O molecule, the results using the basis-set correction are actually worst than the CIPSI ones when using the aug-cc-pVDZ basis set, the error increasing from about 0.003 a.u. for CIPSI to about 0.009 a.u. with the basis-set correction. One should nevertheless keep in mind that the convergence of the dipole moment of H$_2$O is non monotonic at the CCSD(T) level, the dipole moment obtained with the aug-cc-pVDZ basis set being closer to the CBS limit than the ones obtained using the aug-cc-pVTZ or aug-cc-pVQZ basis sets. Therefore, the seemingly good values obtained at the CCSD(T) and CIPSI levels using the aug-cc-pVDZ basis set are likely to be due to a compensation of errors. With the aug-cc-pVTZ basis set, the expected trend is recovered, with CIPSI giving an error of about 0.006 a.u. and the basis-set correction reducing this error to about 0.001 a.u..

\section{Conclusion}
\label{sec:conclusion}

In the present work, we have established the fully self-consistent density-based basis-set correction scheme\cite{GinPraFerAssSavTou-JCP-18}. Differently from previous works where a non-self-consistent approximation was used\cite{GinPraFerAssSavTou-JCP-18,LooPraSceTouGin-JCPL-19,GinSceTouLoo-JCP-19,LooPraSceGinTou-JCTC-20,GinSceLooTou-JCP-20,YaoGinLiTouUmr-JCP-21}, here the energy is minimized in the presence of the basis-set correction functional which i) guarantees to get a lower total energy with respect to the non-self-consistent approximation, and ii) allows the wave function to change under the presence of the basis-set correction. We have tested this scheme on a few atomic and molecular systems (Be, BH, FH, H$_2$O, and CH$_2$) with CIPSI wave functions and two different basis-set correction functionals $\pbeuegXi$ and $\pbeontns$. While $\pbeuegXi$ is a functional of the density, $\pbeontns$ uses in addition the on-top pair density of the wave function as an independent variable.

The main results are that i) the lowering in total energy is extremely small compared to the non-self-consistent approximations (typically less than $1\%$), which thus justifies this approximation for energy-only calculations, and ii) the wave functions obtained from the self-consistent basis-set correction scheme yield dipole moments which converge much faster with respect to the size of the basis set than standard wave-function calculations, being already close to the CBS value with a triple-zeta basis set. This study thus tend to demonstrate that the density-based basis-set correction scheme is not only useful for energy calculations but also for calculations of response properties.

\section*{Acknowledgement}
This project has received funding from the European Research Council (ERC) under the European Union's Horizon 2020 research and innovation programme Grant agreement No. 810367 (EMC2).

 \end{document}